\documentclass[10pt,letterpaper]{article}
\usepackage[top=0.85in,left=2.75in,footskip=0.75in,marginparwidth=2in]{geometry}

\usepackage[utf8]{inputenc}

\usepackage{cite}

\usepackage{nameref,hyperref}

\usepackage[right]{lineno}

\usepackage{microtype}
\DisableLigatures[f]{encoding = *, family = * }

\raggedright
\usepackage{changepage}

\usepackage[aboveskip=1pt,labelfont=bf,labelsep=period,singlelinecheck=off]{caption}

\makeatletter
\renewcommand{\@biblabel}[1]{\quad#1.}
\makeatother

\usepackage{lastpage,fancyhdr,graphicx}
\usepackage{epstopdf}
\fancyheadoffset[L]{2.25in}
\fancyfootoffset[L]{2.25in}

\usepackage{color}

\definecolor{Gray}{gray}{.25}

\usepackage{graphicx}

\usepackage{sidecap}

\usepackage{amsmath,amssymb}

\usepackage{wrapfig}
\usepackage[pscoord]{eso-pic}
\usepackage[fulladjust]{marginnote}
\reversemarginpar

\usepackage[nodayofweek]{datetime}
\newdateformat{mytoday}{\monthname[\THEMONTH] \twodigit{\THEDAY}, \THEYEAR}

\usepackage{multirow}
\usepackage{booktabs}
\usepackage{rotating}
\usepackage{tabulary}
\newcolumntype{K}[1]{>{\centering\arraybackslash}p{#1}}

\begin{document}
\noindent This is a pre-print of an article published in \textit{Cognitive Computation}. The final authenticated version is available online at: \href{https://doi.org/10.1007/s12559-018-9543-3}{ https://doi.org/10.1007/s12559-018-9543-3}

\ \\
\noindent \textcolor{blue}{Cite this article as:}

\ \\
\noindent \textcolor{blue}{M. Mahmud, M.S. Kaiser, M.M. Rahman, M.A. Rahman, A. Shabut, S. Al-Mamun, A. Hussain. (2018). A Brain-Inspired Trust Management Model to Assure Security in a Cloud based IoT Framework for Neuroscience Applications. \textit{Cogn. Comput.}, doi: 10.1007/s12559-018-9543-3.
\ \\
\noindent \textcopyright\ 2018, Springer Nature holds the copyright of this article.
}

\vspace*{0.35in}
\begin{flushleft}
{\Large
\textbf\newline{A Brain-Inspired Trust Management Model to Assure Security in a Cloud based IoT Framework for Neuroscience Applications}
}
\newline
\\
Mufti Mahmud\textsuperscript{1,*},
M. Shamim Kaiser\textsuperscript{2,*},
M. Mostafizur Rahman\textsuperscript{3}, 
M. Arifur Rahman\textsuperscript{4},  
Antesar Shabut\textsuperscript{5},
Shamim Al-Mamun\textsuperscript{6}, 
Amir Hussain\textsuperscript{7}
\\
\bigskip
\textsuperscript{1} NeuroChip Lab, University of Padova, 35131 - Padova, Italy
\\
\textsuperscript{2} IIT, Jahangirnagar University, Savar, 1342 -  Dhaka, Bangladesh
\\
\textsuperscript{3} American International University - Bangladesh, 1213 - Dhaka, Bangladesh
\\
\textsuperscript{4} Department of Computer Science, University of Sheffield, Sheffield, S10 2TN, UK
\\
\textsuperscript{5} Anglia Ruskin University, CM1 1SQ - Chelmsford, UK
\\
\textsuperscript{6} Saitama University, Saitama, 338-8570, Japan
\\
\textsuperscript{7} Division of Computing Science \& Maths, University of Stirling, FK9 4LA Stirling, UK \\
\bigskip
\textsuperscript{*} Co-`first and corresponding' author. 
Emails: muftimahmud@gmail.com (M. Mahmud), mskaiser@juniv.edu (M.S. Kaiser) \\
\bigskip

\end{flushleft}

\section*{Abstract}

Rapid popularity of Internet of Things (IoT) and cloud computing permits neuroscientists to collect multilevel and multichannel brain data to better understand brain functions, diagnose diseases, and devise treatments. 
To ensure secure and reliable data communication between end-to-end (E2E) devices supported by current IoT and cloud infrastructure, trust management is needed at the IoT and user ends. 
This paper introduces a Neuro-Fuzzy based Brain-inspired trust management model (TMM) to secure IoT devices and relay nodes, and to ensure data reliability. The proposed TMM utilizes node behavioral trust and data trust estimated using Adaptive Neuro-Fuzzy Inference System and weighted-additive methods respectively to assess the nodes trustworthiness. 
In contrast to the existing fuzzy based TMMs, the NS2 simulation results confirm the robustness and accuracy of the proposed TMM in identifying malicious nodes in the communication network.
With the growing usage of cloud based IoT frameworks in Neuroscience research, integrating the proposed TMM into the existing infrastructure will assure secure and reliable data communication among the E2E devices.



\section*{Introduction}
\label{sec-intro}

In recent years biological data has grown significantly, thanks to the technological developments, now scientists can acquire data simultaneously from multiple levels and channels of a living system \cite{mahmud_dl_rl_biol_2017}, and simulate large scale brain networks \cite{schadt_cloud_genetics_2010,shahand_neuro_gateway_2015}. One of the major contributors to this biological big data is Neuroscience \cite{landhuis_neuro_bigdata_2017}. Brain signals, e.g., Electroencephalogram (EEG), Electrocorticogram (ECoG), Neuronal Spikes (AP), Local Field Potentials (LFPs) along with brain imaging techniques, e.g., Magnetoencephalography (MEG), Magnetic Resonance Imaging (MRI), Functional MRI (fMRI), Positron Emission Tomography (PET) have been extensively used in diagnosis of neurodegenerative diseases \cite{sakkalis_eeg_cogn_res_2011,mcmillan_neurodegenerative_2016}, neuropsychiatric disorders \cite{liu_neuroimaging_review_2015}, and developmental disorders such as Autism Spectrum Disorder \cite{aljawahiri_autism_2017}. Additionally, this data has been effectively utilized in developing various data-driven disease models \cite{young_data_driven_2014,burns_data_intensive_2014}.

\begin{figure}[!btp]
\centering
\includegraphics[scale=1]{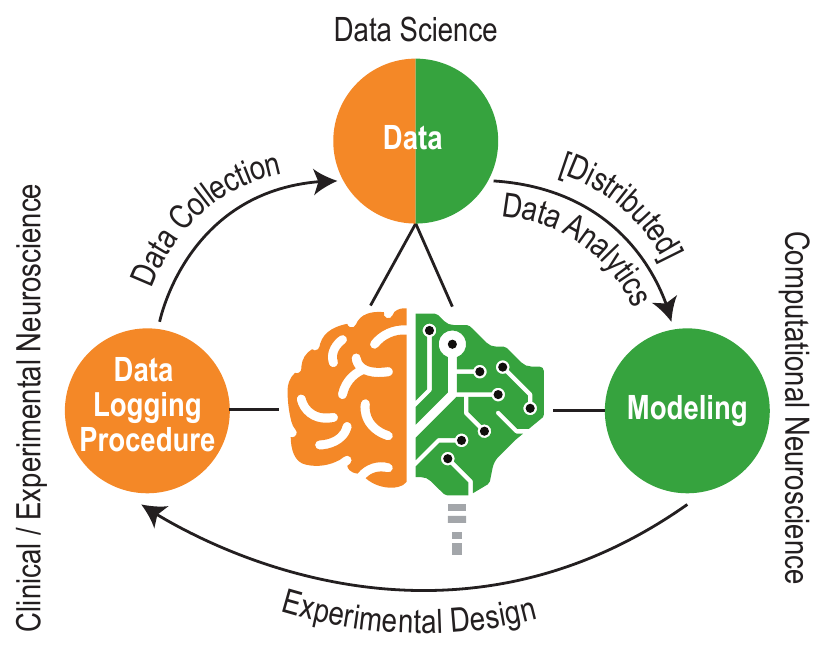}
\caption{Cycle of modern Neuroscience research.} 
\label{fig-concept}
\end{figure}

Modern day Neuroscience research is driven by data (see Fig. \ref{fig-concept}). Both clinical and experimental neuroscience research generate huge amount of data \cite{mahmud_qst_2014} and analyzing those data to draw meaningful conclusions is very challenging \cite{mahmud_sig_proc_review_2016}. The extracted knowledge from these data allow the development and refining of data-intensive models and describe the underlying biological phenomena which in turn facilitate experimental design \cite{ncc_cloud_2016}. 
The data analytics and modeling phases are computationally intensive, and advancements in artificial intelligence \cite{luo_bics_2016} and cloud computing \cite{hashem_bigdata_cloud_review_2015} allowed scientists to perform these steps smoothly.
The `cloudification' greatly facilitated scientists by providing `software as a service' (e.g., service oriented architecture or SOA) instead of running the data-intensive analyses and modeling locally in the computers. In other words, cloud computing and big data paradigms converted context-aware research into exhaustive, data-driven research.

Now, with the emergence of the Internet of Things (IoT), various sensors can be connected to the cloud for seamless resource sharing. Such IoT-Cyber Physical Systems (IoT-CPS) provide a platform to data-driven research and design appropriate medical services for patients. 
The IoT-CPS tailored to patient monitoring and care are around for a few years now and it allowed hospitals and healthcare processionals to 
seamlessly exchange patients' data even from remote locations. These data may represent a wide range of healthcare parameters collected through the IoT for healthcare (IoHT) sensors. One of the main challenges of this type of IoT-CPS is to ensure privacy and information security. Thus, the trust management plays a vital role for the end users which act as a first step of information security.  Despite the fact that trust management is required for all such frameworks dealing with biological data acquirable through the IoHT devices, the Neuroscience data stands apart from the others and requires special attention due to their high variability and spontaneity. While in many biosignals (e.g., Electrocardiogram, Electromyogram) periodicities and similarities have been noticed in terms of frequency content, amplitude and shape, the Neuroscience data (e.g., EEG, ECoG, LFPs, AP, etc.) have been known for their variabilities \cite{mahmud_lfp_sorting_2012,mahmud_sigmate_2012,mahmud_single_lfp_2016} making them more prone to misidentification, misclassification and misinterpretation in cases when the signals are unsupervisedly acquired without any experts. Therefore, to design robust telemedicine  systems using IoT-CPS targeting Neuroscience applications, extra care must be taken to ensure the trustworthiness of the IoHT nodes.

Mahmud et al. introduced a service-oriented architecture for web based collaborative biomedical signal analysis \cite{mahmud_service_2012}. As an initial platform with three main components (i.e., users, contributors, and services), this model assumed the inherent security of the internet and used certificate based security as authentication scheme for the contributors and users to deploy and utilize services. The same architecture can be extended by delegating the data coming from the IoT devices to the cloud for analysis. Additionally, a cloud-based healthcare system was proposed in \cite{Zhang_Health-CPS_2017} to provide convenient patient-centric healthcare services. In this model, the cloud performed the big data analytics and the authors reported significant performance improvement in the cloud-based system which too can be adapted to suit smart healthcare applications. Also, biologically inspired cloud resource provisioning was proposed for optimal handling of big healthcare data \cite{ullah_ss_cloud_2016}. 

While the assumption of a secure cloud is appropriate in the context of currently discussed communication models, discarding malicious transmission -- identified by the nodes profile information, behavior, and data similarity -- is vital to ensure the optimized performance, reliability, and robustness of a system. In the current scenario, profile information is validated by the authentication services, and the nodes behavior and data similarity are handled by a trust management system. To make a more trustworthy system, Shabut et al. identified the malicious nodes based on their behavior and improved packets delivery through a multi-hop relay network excluding those misbehaving nodes \cite{Shabut_Recom_trust_2015}. Another work proposed a dynamic cluster based recommendation model to minimize the data sparsity or cold start situations using nodes behavior to improve quality of service (QoS) of end-to-end (E2E) transmission \cite{Shabut_sparsity_2017}. 
\ 
Chen et al. proposed a Fuzzy reputation-based trust model (TRM) for IoT-CPS which estimated the nodes trust from their behavior and showed an improved performance in comparison to a communication system without trust \cite{ Chen2011TRMIoTAT}. An ant colony based trust model was presented to determine the trust value of wireless nodes which exhibited improved accuracy \cite{marzi_enhanced_2013}. 
Context-aware multiservice trust management systems were proposed in \cite{ben_saied_trust_2013,dolera_tormo_dynamic_2015} which filtered malicious nodes in the E2E and heterogeneous IoT architectures with high accuracy. Another trust management model (TMM) was proposed to evaluate the trustworthiness of nodes in the wireless sensor network through beta distribution. The aggregated trust value from data and energy was used in identifying the untrustworthy relay nodes to reduce the internal threats \cite{fang_btres:_2016}. Yet another trust management system, based on an agent's trustworthiness and confidence, was proposed to evaluate the trustworthiness of the IoT nodes \cite{Ruan_TMF_2016}. Moreover, a joint social and QoS TMM was presented to find the trust level of wireless nodes in a mobile adhoc network \cite{Chen_Social_Qos_trust+2013}. 

However, identifying the malicious transmission using only nodes behavior isn't enough to ensure reliable communication. It is important to guarantee that the data generated by the nodes are error-free -- which is a big challenge -- and a TMM that takes into account both nodes behavior and data similarity can be a solution to confirm nodes reliability. 

This paper presents an Adaptive Neuro-Fuzzy based Brain-inspired TMM targeting cloud based IoT architecture to determine data trust and behavioral trust for all IoT devices and relay nodes to ensure reliable data communication between E2E devices. This work also investigates the effects of trust management on the QoS issues of the cloud based IoT architecture suitable for neuroscience applications.

\begin{figure*}[!bthp]
\hspace{-4cm} 
\includegraphics[scale=1]{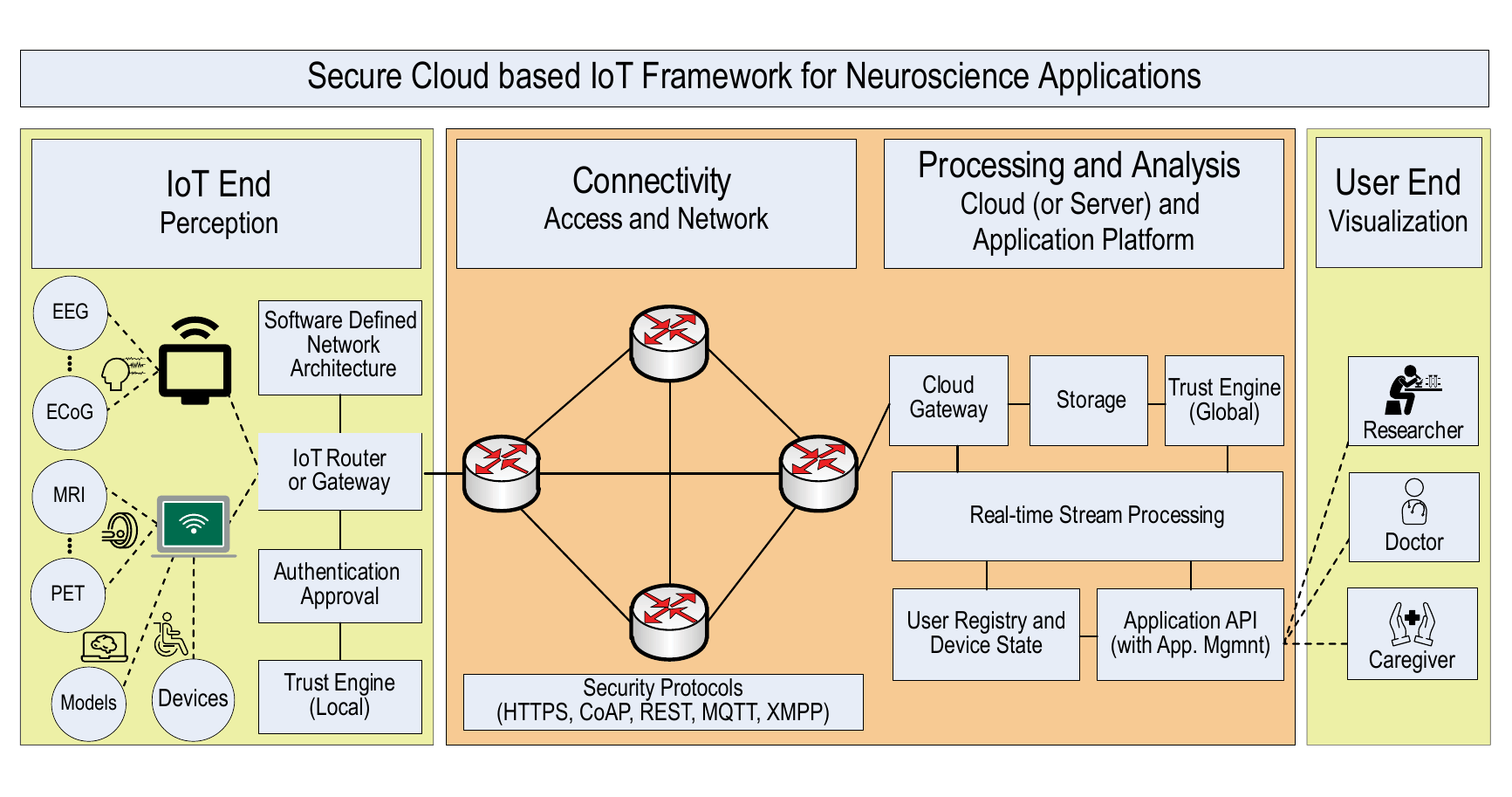}
\caption{Cloud based IoT Architecture for Neuroscience Applications. All the IoT sensor nodes are deployed in the perception layer (IoT site).} 
\label{fig-IoT-Cloud}
\end{figure*}

\section{Cloud based IoT Architecture}
\label{sec:NS}

The big data and cloud are two paramount elements for creating collaborative frameworks to analyze brain signals (e.g., EEG, ECoG, AP, LFPs, etc.) and brain images (e.g., MEG, MRI, fMRI, PET, etc.) and to perform data-driven modeling \cite{mahmud_service_2012}. Due to the wide range of advantages offered by such architectures, they have become the trend  in recent years \cite{Zhang-health-cps_2017}.

\begin{figure}[!btp]
\centering
\includegraphics[scale=1]{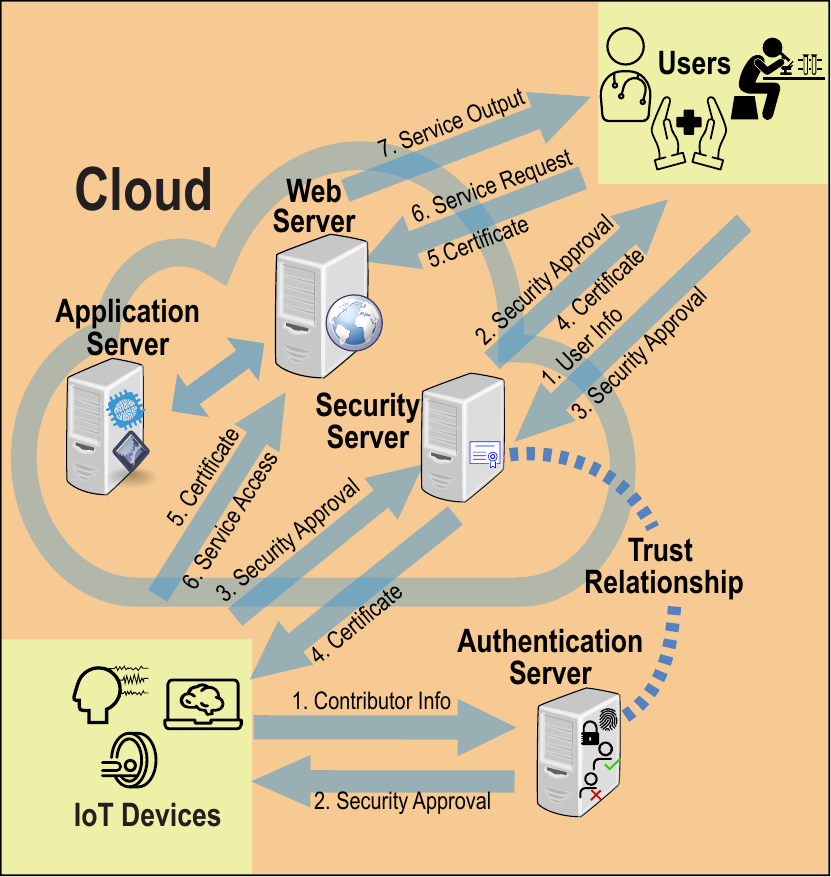}
\caption{Cloud authentication model (adapted from \cite{mahmud_service_2012}).} 
\label{fig-cloudauth}
\end{figure}

Focusing on applications related to Neuroscience, Fig. \ref{fig-IoT-Cloud} illustrates a cloud based IoT framework which consists of three main components, i.e., the IoT end (contains the data generating devices), the cloud component (provides the access and connectivity, and processing and analysis of data), and the user end (provides the analyzed and processed data to the users, e.g., doctors, caregivers, and researchers).
In this framework, the data from various Neurotechnology empowered devices are collected for the development of state-of-the-art techniques pertaining to intelligent healthcare and advancement of Neuroscience research. 
At the IoT end, also known as perception layer, various data generating devices are connected to respective transceiver devices to forward the data to the cloud through the IoT gateway either for data analytics or simply for storage. Additionally, the brain signals generated at the IoT end are also used in operating various medical and assistive devices (e.g., automatic wheelchair, robotic arm, etc.) \cite{kaiser_neuro-fuzzy_2016,Zhang-health-cps_2017} to provide the better monitoring and improve the quality of life.
The cloud is used for defining the access and the network and perform data storage and analytics. Extending the work of Mahmud et al. \cite{mahmud_service_2012}, in our framework, we consider the cloud to be secure through existing certification and authentication models (see Fig. \ref{fig-cloudauth}). 
Finally, at the user end, the service consumers can access and visualize the processed data based on granted rights and privileges.

\begin{figure}[!bthp]
\centering
\includegraphics[scale=1]{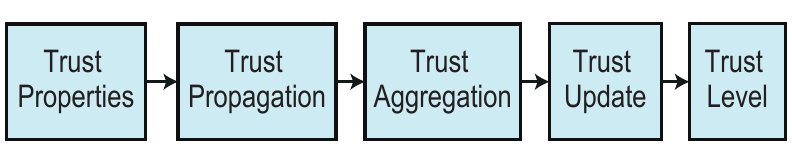}
\caption{Block diagram showing various steps of a trust evaluation process. } 
\label{fig-trust-steps}
\end{figure}

In the cloud based IoT architectures, the IoT devices or nodes generate data owing to various Neuroscience applications. Like human relationships, these nodes collaborate with each other through certain predefined social properties, and these properties are the `Trust Compositions' (see section \ref{sec:Model}). The values of these social properties are propagated on the IoT and user ends (known as `Trust Propagation'). During direct or indirect interactions, the trust metrics of each node are aggregated through static weighted sum, neuro-fuzzy method, and Bayesian inference (known as `Trust Aggregation'). The trust value of each node is then updated when an interaction is completed (known as `Trust Update'). This update can also be done periodically for energy efficiency. The block diagram of the trust management steps is illustrated in Fig. \ref{fig-trust-steps}.

\begin{figure}[!btp]
\centering
\includegraphics[scale=1]{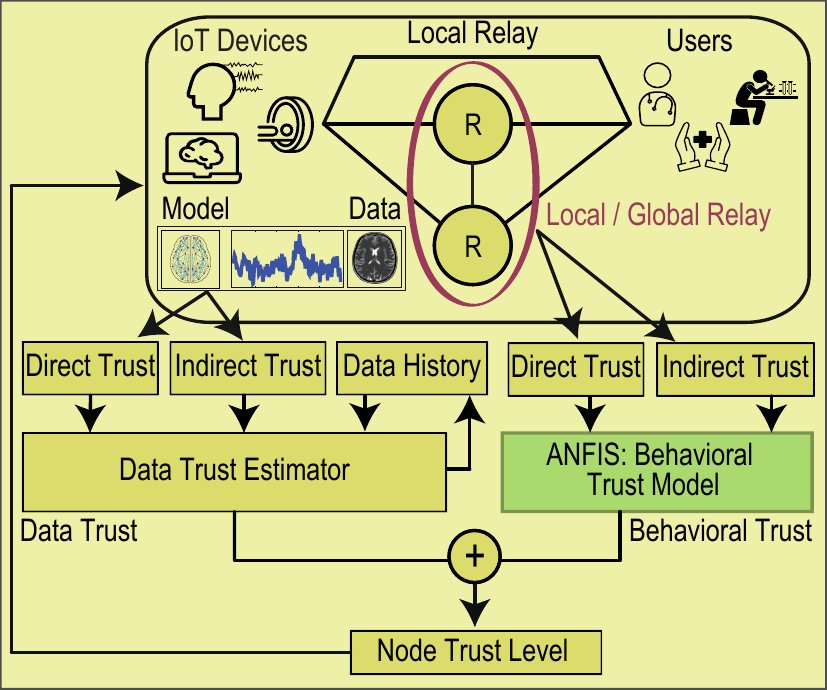}
\caption{The trust management model. Data trust and behavioral trust values are aggregated to find the trust level of the sensor and relay nodes.} 
\label{fig-model}
\end{figure}

\section{Trust Management Model}
\label{sec:Model} 

The proposed TMM is illustrated in Fig. \ref{fig-model}, where the IoT nodes directly or via local/global relay nodes (such as smartphones, routers, etc.) interact with the sensor hubs (see Fig. \ref{fig-IoT-Cloud}) to establish successful communication links. The individual trust levels of the IoT devices and relay nodes are required to be evaluated to discard the malicious nodes \cite{yan_iot_trust_review_2014}.

As the data communication in the access and cloud layer is secured, the IoT and user ends are the main focus of our TMM for ensuring the E2E trust among IoT devices and users for cloud based Neuroscience applications. Mimicking the social relation of people, the IoT devices and relay nodes are assumed to have social relationships among themselves. Thus, the interactions and collaborations among these nodes are employed to evaluate the trust level of each node. In deducing E2E trust level, certain relationship among the nodes are considered which include-- node profile information, node behavioral trust, and data trust  \cite{afsana_trust_energy_2015}. 

The profile information is assured by the authentication service, whereas, the latter two are estimated using adaptive neuro-fuzzy inference system (ANFIS) and weighted-additive method, respectively.
The node behavioral evidence is assessed through direct and indirect interactions among the nodes. For each node, the assessment of the behavioral trust is performed considering three factors related to that node--  relative frequency of interaction (RFI), intimacy, and honesty. 
The data trust is assessed by estimating the deviation of a node's instantaneous data from the historical data of that node. Both direct and indirect methods can be employed to evaluate data trust of a node.  

Mathematically, the trust level of a given node ($j$) denoted by $\mathcal{T}_{j}$ is estimated by summing up the behavioral and data trust as Equation \ref{eq:node}.

\begin{equation} 
\label{eq:node}
\centering
\mathcal{T}_{j}(t)=\mathcal{T}_{j}^{nb}(t)+\mathcal{T}_{j}^{d}(t),
\end{equation}

\noindent
where, $\mathcal{T}_{j}^{nb}(t)$ is the evaluated behavioral trust and $\mathcal{T}_{j}^{d}(t)$ is the evaluated data trust.

\subsection{Evaluating Behavioral Trust}
\subsubsection{Behavioral Trust Metrics}
\label{subsubsec:trust} 
The trust properties for the behavioral trust of a nodes are discussed below.

\noindent
\paragraph{Relative Frequency of Interaction (RFI).}
Zhang et al. studied the interaction frequency among nodes \cite{zhang_effects_2001}. 
The interaction frequency refers to the number of interactions, between the assessor and assessee, that take place within a given unit of observation time. The higher the successful interaction rate, the higher the degree of closeness. It means the assessee node is a trustworthy node.
It has also been reported that the closeness in a relationship (e.g., friendship) can be predicted from the past interaction and it confound the future interaction \cite{josang_survey_2007,cherry_entrepreneur_2014}. Therefore, the RFI-aware trust, $\mathcal{T}_{j}^{RFI}$, can be calculated by Equation \ref{eq:fi}.
\begin{equation}\label{eq:fi}
\mathcal{T}_{j}^{RFI}=\frac{n_j}{N},
\end{equation}
\noindent
where $n_j$ is the number of interactions between the assessee node $j$ and the assessor node in an observation period $t$,
whereas, $N$ is total number of interactions between node $j$ with other $k$ nodes during $t$. 

\noindent
\paragraph{Intimacy.}
In any social context, the intimacy or relationship duration of interaction is an important factor in calculating the trust level. The higher is the time of interaction between an assessee node and an assessor or guarantor node, the higher is the intimacy. Considering the total time spend of an assessor node $i$ with the assessee node $j$ as $t_{ij}$ and the cumulative time spend of $j$ with other $k$ guarantor nodes as $t_{kj}$, the intimacy ($\mathcal{T}_{j}^{I}$) can be calculated by Equation \ref{eq:int} \cite{daly_social_2009}. 
\begin{equation} \label{eq:int}
\mathcal{T}_{j}^{I}=\frac{t_{ij}}{t_{ij}-t_{kj}}.
\end{equation}

\noindent
\paragraph{Honesty.}
Honesty is one of the main factors for establishing social trust between two given nodes. It can be determined using the successful and unsuccessful interactions of those nodes. 
Usually, the value of honesty lies between [0,1], i.e., $\mathcal{T}_{j}^{H}\in [0,1]$. In other words, $\mathcal{T}_{j}^{H}=0$ means no successful interaction, and $\mathcal{T}_j^{H}(t)\rightarrow 1$ means the assessee node $j$ is a trustworthy node. 
While $a_j$ and $b_j$ denote successful and unsuccessful interactions respectively, their values are estimated using the Beta distribution \cite{momani_beta_distr_2014,liu_reputation_2017}, where the distribution $f(p|a_j,b_j)$ is expressed by the Gamma function $\Gamma(\cdot)$ with $0\leq p \leq 1$, $a_j>0$, $b_j>0$; and $p\neq0$ if $a_j<1$ and $p\neq1$ if $b_j<1$ \cite{josang_beta_2002}.
Finally, the honesty aware trust value can be calculated by Equation \ref{eq:hon}.
\begin{equation} \label{eq:hon}
    \mathcal{T}_{j}^{H}(t)=\frac{a_j}{a_j+b_j}.
\end{equation}

\subsubsection{Node Behavioral Trust}
\label{subsubsec-nbt}
The node behavioral trust is calculated from both direct and indirect interactions between nodes. At a given time $t$, an assessor node directly interacts with the assessed node and evaluates the direct trust level (i.e., $\mathcal{T}_{j}^{d,nb}(t)$) from the previous direct interactions. Based on the guarantee provided by the adjacent nodes the indirect trust level (i.e., $\mathcal{T}_{kj}^{ind,nb}(t)$) can be evaluated. The guarantor nodes ($k$ number of nodes) provide guarantee based on the previous interactions with the assessed node. The behavioral trust of $j$-th node is given by Equation \ref{eq:nodebehavioral}.
\begin{equation} \label{eq:nodebehavioral}
\mathcal{T}_{j}^{nb}(t)=\mathcal{T}_{j}^{d,nb}(t)+\sum_k \frac{1}{\mathcal{H}_k}\mathcal{T}_{kj}^{ind,nb}(t),
\end{equation}
\noindent
where $\mathcal{H}_k$ is the hop count for the $k$-th guarantor node.

\subsubsection{ANFIS based Node Behavioral Trust Model}
\label{subsubsec:NFtrust}
Fuzzy inference system (FIS) is a rule based expert system which can mimic Brain's logical inference to represent a system. In ANFIS, a fuzzy inference system is employed to represent a nonlinear system with any complexity. The parameters of the input and output membership functions can be tuned by the backpropagation or hybrid backpropagation-least squares algorithm \cite{Takagi-FIS-1985,Hmouz_ANFIS_2012}. Due to its adaptive nature, the ANFIS is more powerful in comparison to FIS.

\begin{figure}[!bthp]
\centering
\includegraphics[scale=1]{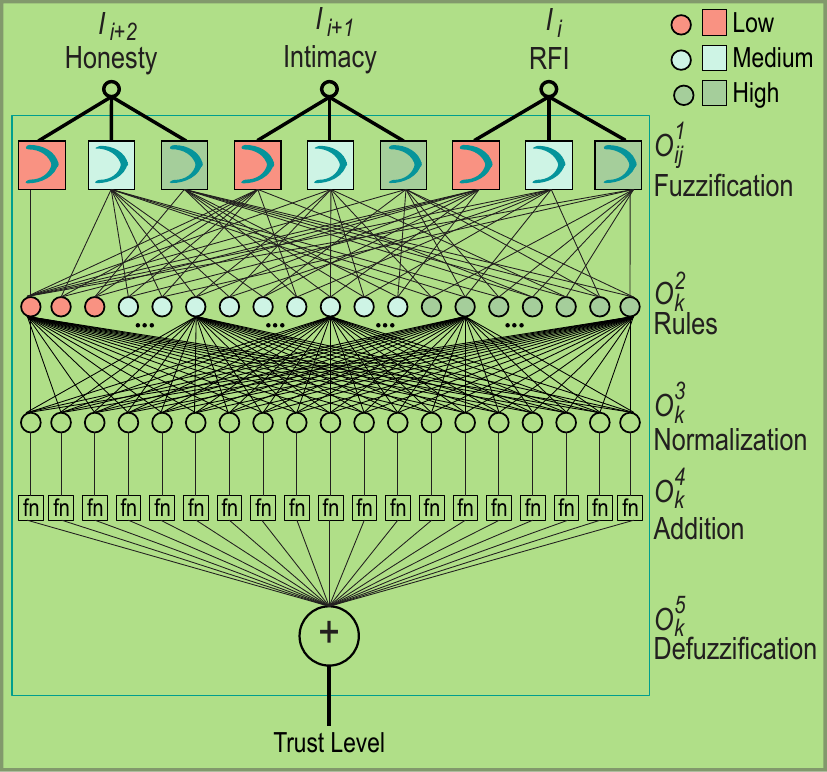}
\caption{ANFIS for the node behavioral trust calculation. The model evaluates node behavioral trust based on the RFI, Intimacy, and Honesty. The `fn' denotes the $y_k$ function in the form $y_k=\sum_i w_{ki}I_i+b_k$.} 
\label{fig-anfis-model}
\end{figure}

The node behavior is evaluated by the ANFIS model as illustrated in Fig. \ref{fig-anfis-model}. 
The system consists of three inputs --relative frequency of interactions (RFI), Intimacy, and Honesty. Each input has three linguistic terms or membership functions (MFs), i.e., \textit{Low}, \textit{Medium}, and \textit{High}. Therefore, there are nineteen possible IF-THEN rules in the rule based system (see Fig. \ref{fig-anfis-model}) and one output called node behavioral trust level.

There are five layers-- Fuzzification, Rule, Normalization, Defuzzification and Output. 
Detailed description of each of these layer is described in \cite{Takagi-FIS-1985,Hmouz_ANFIS_2012,kaiser_neuro-fuzzy_2016}. The outputs of the layers are expressed by: 
\begin{align*} 
\textrm{Fuzzification: \;} O_{ij}^1&=\mu_{ij}(I_i), \\ 
\textrm{Rule: \;} O_{k}^2&=\prod O_{ij}^1=\prod \mu_{ik}(I_i), \\ 
\textrm{Normalization: \;} O_{k}^3&=\frac{ O_{k}^2}{\sum_k O_{k}^2}, \\ 
\textrm{Defuzzification: \;} O_{k}^4&= O_{k}^3 y_k, \; y_k=\sum_i w_{ki}I_i+b_k, \\ 
\textrm{Output: \;} O_{k}^5&=\mathcal{T}_{j}^{nb}(t)=\sum_k O_{k}^4, 
\end{align*}
where, $i=1,2,3$; $j=1,2,3$; $k=1,2,...,19$; $\mu_{ij}$ is $j$-th MF for input $I_i$,  
$w_ki$ and $b_k$ are consequent parameters; and $\mathcal{T}_{j}^{nb}(t)$ is the behavioral trust level of $j$-th node. 

\begin{figure}[!bthp]
\centering
\includegraphics[scale=1]{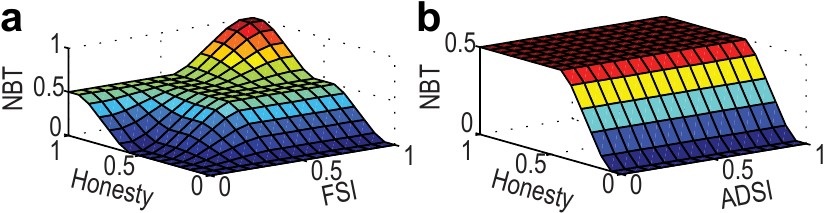}
\caption{The output surface plots of ANFIS where node behavioral trust is plotted against the trust properties (a) Honesty and RFI, and (b) Honesty and Intimacy.}
\label{fig-surfaceplot}
\end{figure}

The ANFIS model is trained with the input-output datasets generated from the NS2 simulator \cite{issariyakul_ns2_2009}.
This dataset is generated for the placement of 50 nodes where a percentage of the nodes are configured as misbehaving nodes. 
Beta distribution calculated the failure and success of the interactions. 
For the predefined rule-based, the ANFIS model has changed the MFs, and premise/ consequent parameters for finding the node-behavior trust value. 
Fig. \ref{fig-surfaceplot} shows the output surface plots of ANFIS model where node behavioral trust is plotted against the trust properties (a) Honesty and RFI, and (b) Honesty and Intimacy.

\subsection{Evaluation of Data Trust}
The data trust of a node consists of direct and indirect trust based on the historical data of the node(s).

\noindent
\paragraph{Direct Data Trust.}
The value of direct data trust depends on the deviation of a node's instantaneous data from its historical data. The historical data are the average value of the node's data for a recent period ($T$). Mathematically, the direct data trust, $T_{j}^{dd}(t)$, of the $j$-th node with the $i$-th relay can be expressed by equation  \ref{eq:ddt}
\begin{equation} \label{eq:ddt}
   T_{j}^{dd}(t)=\left\{
                \begin{array}{ll}
                  T_{max} \; \; \textrm{for \;} D_{j}^{dd}(t)=D^{his}\\
                  \frac{1}{|D_{j}^{dd}(t)-D^{his}|}  \; \; \textrm{for \;} D_{j}^{dd}(t) \neq D^{his},                   \end{array}
              \right.
\end{equation}
\noindent
where, $D_{j}^{dd}$ is the instantaneous data of $j$-th node during direct interaction whereas $D^{his}$ is the historical data.

\noindent
\paragraph{Indirect Data Trust.}
The indirect data trust, $T_{kj}^{di}$ is the average value of the deviation of a node's instantaneous data from the historical data of $k$ nodes with $j$-th relay under the assumption that the included nodes are all trusted. Mathematically, $T_{j}^{di}(t)$ can be expressed by the equation  \ref{eq:didt}
\begin{equation} \label{eq:didt}
T_{j}^{di}(t)=\left\{
                \begin{array}{ll}
                  T_{max} \; \; \textrm{for \;} \frac{\sum_k D_{kj}^{ind}(t)}{k}=D^{his}\\
                  \frac{1}{|\frac{\sum_k D_{kj}^{ind}(t)}{k}-D_{j}^{his}|}  \; \; \textrm{for \;} \frac{\sum_k D_{kj}^{ind}(t)}{k} \neq D_{j}^{his},                   \end{array}
              \right.
\end{equation}
\noindent
where, $D_{kj}^{ind}$ is the instantaneous data of $j$-th node during indirect interaction with $k$ nodes.

Having obtained the direct and indirect trust values, data trust of the $j$-th node is calculated by Equation \ref{eq:datatrust}
\begin{equation} \label{eq:datatrust}
\mathcal{T}_{j}^{d}(t)=\mathcal{T}_{j}^{dd}(t)+\sum_k \frac{1}{\mathcal{H}_k}\mathcal{T}_{kj}^{di}(t-t_m),
\end{equation}
\noindent
where $t_m$ is the $m$-th time.

\section{Performance Metrics}
\label{sec:permetric}
The proposed Brain-inspired TMM, suitable for cloud based IoT frameworks targeting Neuroscience applications, has been evaluated using Packet Forwarding Ratio (PFR) \cite{gopinath_energy_2015};  Network Throughput (NetT) \cite{Kaur_MANET_2015, Gupta_AODV_2013, Talreja_TVF_2016,dhananjayan_t2ar:_2016}; Average Energy Consumption Ratio (AECR) \cite{Ruan_TMF_2016}; Accuracy \cite{kaiser_neuro-fuzzy_2016}; and F-measure \cite{ghosh_novel_2014}.

\paragraph{PFR.} 
The PFR is the ratio between a number of packets received by the IoT CPS and the number of packets transmitted by the source node. The PFR decreases when the forwarded packets are dropped due to reasons like-- buffer overflow, blocking, route failure. Mathematically, the E2E PFR is calculated by Equation \ref{eq:R1}.
\begin{equation} \label{eq:R1}
\textrm{PFR}=\frac{\sum_k PKT_{rec}}{\sum_n PKT_{send}},
\end{equation}
where, $PKT_{rec}$ and $PKT_{send}$ are the number of packets received by the destination node and packets send by the source node. The source node sends $n$ number of packets and destination node receives $k$ number of packets, and $k<n$.

\paragraph{NetT.}
The NetT can be defined as the rate at which the source transmissions are delivered successfully to the destination over the link(s) between the source-destination pair.
The value of the throughput declines with the appearance of misbehaving nodes in the network. Mathematically, the NetT is calculated by equation  \ref{eq:R2}.
\begin{equation} \label{eq:R2}
\textrm{NetT}=\frac{N_{success}}{t_{trans}},
\end{equation}
where, $N_{success}$ is the number of successful transmission delivered to the destination and $t_{trans}$ is the considered transmission interval.

\paragraph{AECR.}
The AECR is an another performance metric which is the ratio between the energy consumption for evaluating a trust metric ($E_{te}$) and the energy consumption for the data transmission (for sending ($E_{send}$) and for receiving ($E_{rec}$)) of a node.
The AECR of a malicious node is lower than that of a legitimate node as a malicious node does not participate in the packet forwarding or route discovery. 
Mathematically, AECR is calculated by Equation \ref{eq:R3}.
\begin{equation} \label{eq:R3}
\textrm{AECR}=\frac{\sum_n E_{te}}{\sum_n(E_{rec}+E_{send})}.
\end{equation}

\paragraph{Accuracy.}
Accuracy is the ratio between the numbers of total successful interactions and total interactions. Mathematically, accuracy $A$ is expressed by Equation \ref{eq:accuracy} \cite{gu_evaluation_2009}.
\begin{equation} \label{eq:accuracy}
A=\frac{TP+TN}{TP+FP+TN+FN},
\end{equation}
\noindent
where, $TP$ is the number of successful interactions categorized as successful, $TN$ is the number of successful interactions categorized as unsuccessful, $FP$ is the number of unsuccessful interactions categorized as successful, and $FN$ is the number of unsuccessful interactions categorized as unsuccessful. 

\paragraph{F-measure.}
The Precision (=$TP/(TP+FP)$) as well as recall (=$TP/(TP+TN)$) are two important measures considered in evaluating a classification outcome \cite{ghosh_novel_2014}. It is calculated by the harmonic mean of both recall and precision, and mathematically it is expressed by Equation \ref{eq:fmeasure}.
\begin{equation} \label{eq:fmeasure}
\textrm{F-measure}=\frac{2}{1/recall+1/precision}.
\end{equation}

\section{Results}
\label{sec:Results} 
To verify the efficacy of the proposed TMM, simulation was performed in the NS-2 platform \cite{issariyakul_ns2_2009}. The parameters and setting employed in this platform are listed in Table \ref{tab:sim}. The results were obtained by running the simulation for twenty times and then taking the average values of these twenty runs. It was assumed that the nodes had wireless capabilities and were communicating either directly or through multihop relay nodes to the IoT-CPS. The Adhoc On-demand Distance Vector (AODV) routing protocol \cite{andel_adaptive_2008} was employed to simulate the communication scenario. The IoT devices or relay nodes were categorized in two types-- legitimate node and malicious node. The legitimate nodes took part in the route discovery and packet forwarding process, whereas the malicious nodes in neither took part in packet forwarding nor in route discovery. 

The ANFIS based TMM was incorporated in the IoT-CPS network and all the nodes were initialized with random trust values. After a certain number of interactions the node behavior trust, and direct and indirect data trust were evaluated by the model.

\begin{table}[!tbhp]
\centering
\caption{Parameters and settings used in simulation.}
\label{tab:sim}       
\begin{tabular}{ll}
\hline\noalign{\smallskip}
Parameters & Numerical Value   \\
\hline\noalign{\smallskip}
Simulator & NS-2  \\
Routing & AODV \\
Node distribution & Random  \\
Traffic & CBR   \\
Nodes & 50   \\
MAC & 802.11   \\
Speed & 3 m/s   \\
Packet size & 512 bytes   \\
Range & 250 m   \\
Max. Connection  & 12   \\
Reply delay  & 60 ms   \\
\noalign{\smallskip}\hline
\end{tabular}
\end{table}

\begin{figure}[!bthp]
\centering
\includegraphics[scale=1]{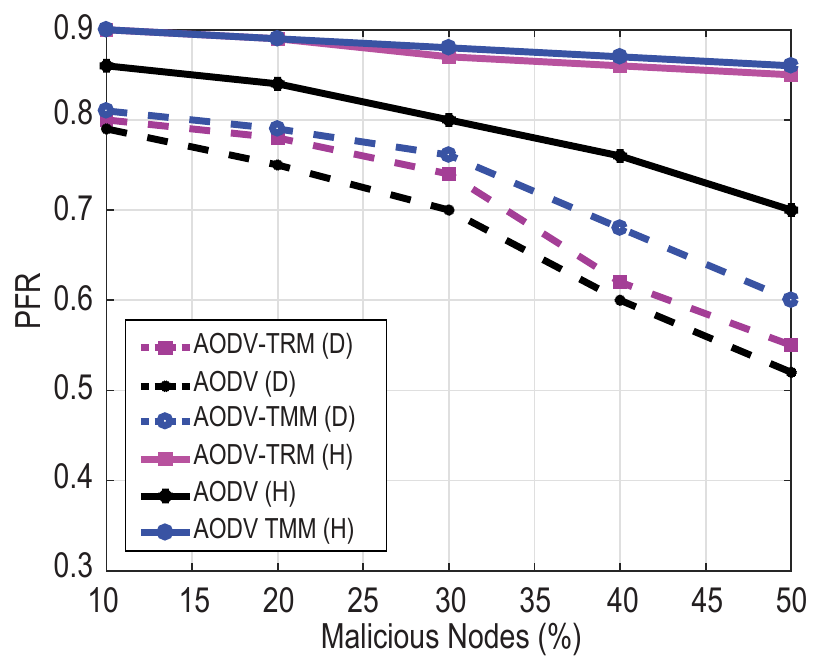}
\caption{The effect of malicious nodes on PFR.} 
\label{fig:NFTM_PFR}
\end{figure}

The PFR dropped significantly when the malicious nodes arose in the IoT or user end. A node was termed malicious if it hid (H) in the route discovery phase or dropped (D) packets intentionally. Fig. \ref{fig:NFTM_PFR} depicts the effect of malicious nodes on the PFR. The PFR decreased as the percentage of malicious nodes increased from 10\% to 50\%. In both cases of malicious behavior, the proposed TMM outperformed TRM \cite{Chen2011TRMIoTAT}. In addition, in terms of PFR, both TMM and TRM achieved better performance compared to AODV with no trust management framework (indicated as `AODV').

\begin{figure}[!bthp]
\centering
\includegraphics[scale=1]{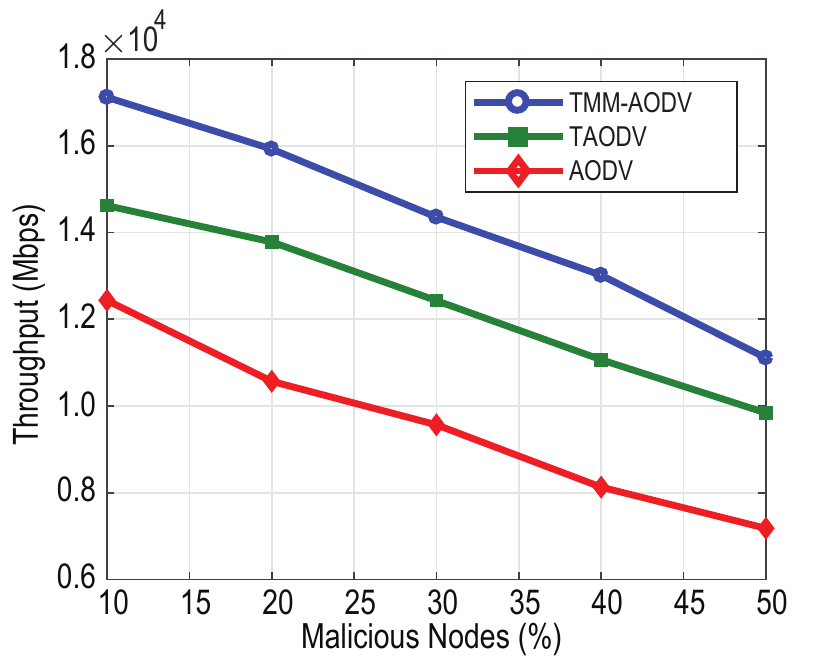}
\caption{The effect of malicious nodes on overall network performance.} 
\label{fig:sim_throughput}
\end{figure}

The malicious nodes changed the overall network throughput as illustrated in Fig. \ref{fig:sim_throughput}. When the number of malicious nodes were increased (10\% to 50\%) and the remaining nodes showed legitimate behavior, the throughput of the network decreased. The performance drop was due to the fact that the appearance of the malicious nodes dropped the packet forwarding in the network. The performance of the proposed TMM (AODV-TMM in Fig. \ref{fig:sim_throughput}) was compared with the trusted AODV (TAODV in Fig. \ref{fig:sim_throughput}) and AODV without trust (AODV in Fig. \ref{fig:sim_throughput}). The results showed that the proposed TMM outperforms the TAODV and AODV.

\begin{figure}[!bthp]
\centering
\includegraphics[scale=1]{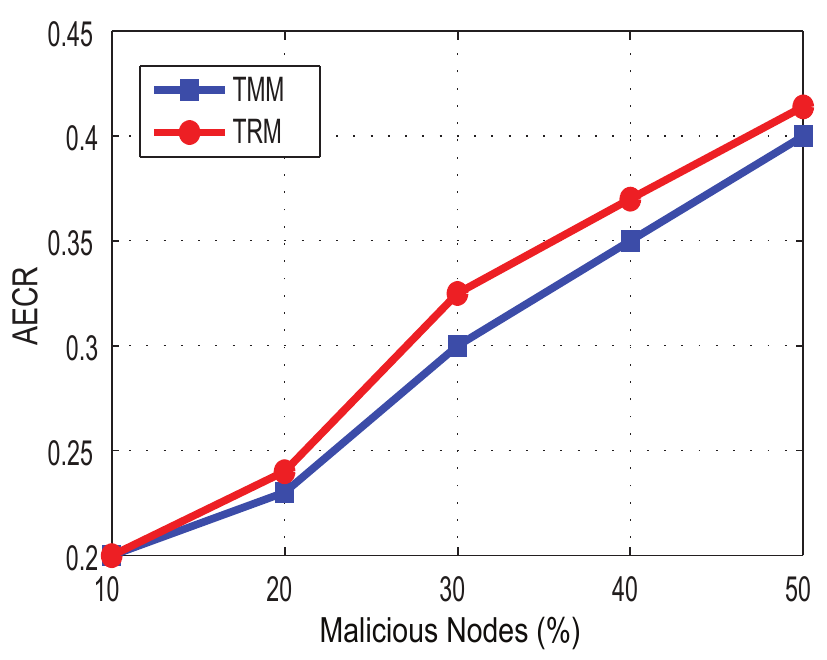}
\caption{The effect of malicious nodes on AECR.} 
\label{fig:sim_aecr}
\end{figure}

Additionally, the proposed TMM is more energy efficient (see Fig. \ref{fig:sim_aecr}). In comparison to the TRM, with the increasing number of malicious nodes (10\% to 50\%) present in the communication network, the proposed TMM consumes less energy during the data transmission process. The reduced AECR value, compared to the TRM, indicates that the proposed TMM is capable of identifying more malicious nodes in the communication network.

Table \ref{tab:performance} shows that the proposed TMM has higher accuracy (0.967 in case 1, when 5 linguistic terms were used: \textit{Very Low}, \textit{Low}, \textit{Medium}, \textit{High}, and \textit{Very High}; and 0.957 in case 2, when 3 linguistic terms were used: \textit{Low}, \textit{Medium}, and \textit{High}) in comparison to a Fuzzy Inference System (FIS) which has an accuracy of 0.89. In addition, the F-measure of the proposed TMM (case 1: 0.97 and case 2: 0.96) also obtained higher values than FIS (0.90).

\begin{table}
\centering
\caption{Performance comparison of three types of Trust management techniques}
\label{tab:performance}       
\begin{tabular}{lcc}
\hline\noalign{\smallskip}
Technique & Accuracy& f-measure   \\
\hline\noalign{\smallskip}
ANFIS (Case 1)	&0.967		&0.97\\
ANFIS (Case 2)	&0.957		&0.96\\
FIS	&0.89		&0.90
\\
\noalign{\smallskip}\hline
\end{tabular}
\end{table}

\section{Conclusion and Future Work}
\label{sec:Conclusion} 

With the unprecedented growth of Brain data and IoT, cloud based data analytics solutions are gaining popularity and now security is a big concern. This paper proposed a Brain-inspired TMM to secure data transmission and ensure data reliability for the cloud-based IoT architecture targeting Neuroscience applications. The TMM evaluates jointly node behavioral trust and data trust using an ANFIS based node behavioral model and a weighted-additive method, respectively. Based on the evaluated trust levels, the model constructs a list of trustworthy nodes.
The performance of the proposed TMM was evaluated regarding PFR, throughput, AECR and accuracy. The NS2 simulation results show that the model performs better than FIS, NFTM and other TM algorithms. In the future, sophisticated optimization techniques along with Bayesian statistics, Deep Learning, and Reinforcement Learning based TMM will be used in ensuring security, reliability and accuracy of the ever growing cloud based IoT and Block Chain architectures.

\ \\
\noindent \textbf{Acknowledgments:} The work was supported by ACS Lab (http://www.acslab.info). Also, the authors express their gratitudes to the members of the ACS Lab for proof-reading the manuscript.

\noindent \textbf{Conflict of Interest Statement}:
The authors declare that the research was conducted in the absence of any commercial or financial relationships that could be construed as a potential conflict of interest.

\noindent \textbf{Authors and Contributors}:
This work was carried out in close collaboration between all co-authors. MM, MSK, MMR, MRA, and SAM first defined the research theme and contributed an early design of the system. MSK and AS further implemented and refined the system development. MM and MSK first drafted the paper and all authors edited the draft. All authors have contributed to, seen, and approved the final manuscript.

\noindent \textbf{Ethical Approval}:
This article does not contain any studies with human participants or animals performed by any of the authors.

\noindent \textbf{Informed Consent}:
As this article does not contain any studies with human participants or animals performed by any of the authors, the informed consent in not applicable.

\bibliographystyle{vancouver} 

\end{document}